\documentclass[11pt,twoside,dvips]{article}
\usepackage{verbatim}
\usepackage{amssymb}
\usepackage{amsfonts}
\usepackage{amsmath}
\usepackage{pifont}
\usepackage[spanish,english]{babel}
\usepackage[pdftex]{graphicx}
\usepackage{verbatim}
\usepackage{epsfig}
\usepackage{bm}
\usepackage{longtable}
\usepackage{fancyhdr}
\begin{document}

\fancypagestyle{plain}{%
\fancyhf{}%
\fancyhead[LO, RE]{XXXVIII International Symposium on Physics in Collision, \\ Bogot\'a, Colombia, 11-15 september 2018}}

\fancyhead{}%
\fancyhead[LO, RE]{XXXVIII International Symposium on Physics in Collision, \\ Bogot\'a, Colombia, 11-15 september 2018}

\title{See-Saw Mechanism in the 2HDM through the simultaneous introduction of a singlet and triplet of Majorana}
\author{Julian Steven Gutierrez$\thanks{%
e-mail: jsgutierrezs@unal.edu.co}$, Carlos Jose Quimbay$\thanks{%
e-mail: cjquimbayh@unal.edu.co}$, \\ Departamento de F\'{i}sica, Universidad nacional de Colombia, \\ Carrera 30 Calle 45-03, CP 111321, Bogot\'{a}, Colombia.∗}
%EndAName
\date{}
\maketitle

\begin{abstract}
\noindent We consider the minimum extension of the SM denominated 2HDM-III, in which,it new coupling constants to the new Higgs field in the Lagrangian of Yukawa are added furtheremore a Majorana mass in generated via See-Saw mechanism SSM-I thus,  the effective masses of the Dirac and Majorana neutrinos will be sensitive to te VEV of Higgs doublets and
to the structure of the Yukawa constants. Then we consider the MSS-III,in which a $Z_{2}$ symmetry is imposed to Lagrangian 2HDM-III to attach a doublet to the particles of the SM, and the other doublet to the triplet of fermions included in this mechanism. Finally, the MSS-I+III is simultaneously implemented  in the M2DH-III. This is important because the additional heavy fermions in the  adjoint representation  of the weak isoespin group $SU(2)$ allows them to interact with the gauge bosons, facilitating their production in the colliding experiments.

\end{abstract}

\section{Introduction}

\noindent Since massive neutrinos require physics beyond the standard model (SM) various models have been proposed with symmetries that rotate the particles and the Higgs fields by spontaneous symmetries breaking  generating the masses for these fermions. One of these models with small extensions of SM correspond to the two Higgs doublets model (2HDM) \cite{1,2}. The presence of two Higgs doublets allows an explanation of the small mass of the neutrinos without reducing Yukawa couplings to extremely small values.

\noindent The see-saw mechanism (SSM) \cite{3} is the most accepted and important scenario to neutrino mass generation nowadays, it shows, first of all, that the massive states are Majorana particles. Secondly, it explains that the small mass of neutrinos when considering a violation of the leptonic number at high energies which is a condition that must be included for Majorana mass term. The explanation of this facts is that there are non-renormalizable terms that can generate very small neutrino masses. The SSM has a new scale of mass that could be associated with the scale of the theory of the great unification (GUT) for the right chirality while for the active neutrinos their mass term is inversely proportional to the mass of the heavy particle. Therefore, the neutrinos become very light.

%\noindent El mecanismo see-saw (MSS) [3], es el escenario m\'{a}s aceptado e importante de la generaci\'{o}n de masa de neutrinos en la actualidad, el cual muestra, en primer lugar, que los estados masivos son part\'{i}culas de Majorana y ademas, da una raz\'{o}n de la peque\~{n}ez de la masa de los neutrinos al considerar una 
%violaci\'{o}n del numero lept\'{o}nico a altas energ\'{i}as, condicion que debe tenerse para t\'{e}rminos de masa de Majorana. Tal explicaci\'{o}n se
%puede dar, b\'{a}sicamente, al decir que existen t\'{e}rminos no-renormalizables que pueden generar masas de neutrinos muy peque\~{n}as. El MSS, tiene una nueva escala de masa que podr\'{i}a estar asociada con la escala de la teor\'{i}a de la gran unificaci\'{o}n GUT, o en general con estas part\'{i}culas respecto a las del MEE, estas part\'{i}culas adicionales, se puede integrar,dando un t\'{e}rmino de masa de Majorana de dimensi\'{o}n 5 para los neutrinos, este t\'{e}rmino de masa es inversamente proporcional a la masa de la part\'{i}cula pesada y por lo tanto, los neutrinos se vuelve muy ligueros.\medskip

\section{Yukawa lagrangian in 2HDM}

\noindent We propose the Lagrangian to induce the breaking of the symmetry $SU(2)_{L}\otimes U(1)_{Y}\rightarrow U(1)_{Q}$ and that is also responsible for the mass of the leptons and the quarks. For the leptonic part, the general lagrangian takes the next structure,
%-------------------------------------------------------------------------------------------------------------------------------------------------
\begin{eqnarray}
-\mathcal{L}_{Y}^{l}=L_{Y}^{III-l}+L_{Y}^{Ext-l} 
\end{eqnarray}
%-------------------------------------------------------------------------------------------------------------------------------------------------
\noindent where $\mathcal{L}^{l}$ represents the Yukawa lagrangian in the 2HDM\footnote{which leaves the mass of the fermions when the Higgs doublets
they acquire a VEV} generalized in the leptonic sector; $L_{Y}^{III-l}$ represents the Yukawa lagranian of 2HDM type-III without right chirality for neutrinos, and $L_{Y}^{Ext-l}$, represents an extension of the Lagrangian, where we have assumed that
 $N_{Rj}=\nu_{jR}$, which originates the mass term for the Dirac neutrinos, as follows.
%---------------------------------------------------------------------------------------------------------------------------------------------------
\begin{eqnarray}
-\mathcal{L}_{Y}^{III-l}&=&\overline{L_{Li}^{0}}\eta _{ij}^{E}\Phi _{1}E_{Rj}^{0}+%
\overline{Q_{Li}^{0}}\eta _{ij}^{U}\widetilde{\Phi }_{1}U_{Rj}^{0}+\overline{%
Q_{Li}^{0}}\eta _{ij}^{D}\Phi _{1}D_{Rj}^{0}  \label{Lagrangiano Yukawa M2DH} 
\\
&&+\overline{L_{Li}^{0}}\xi _{ij}^{E}\Phi _{2}E_{Rj}^{0}+\overline{%
Q_{Li}^{0}}\xi _{ij}^{U}\widetilde{\Phi }_{2}U_{Rj}^{0}+\overline{Q_{Li}^{0}%
}\xi _{ij}^{D}\Phi _{2}D_{Rj}^{0}+h.c.,  \notag
\end{eqnarray}
%---------------------------------------------------------------------------------------------------------------------------------------------------
\noindent Where $i=1,2,3$ is the generation index. The left chirality of the leptons is represented by a doublet $L_{Li}=\left( 
\begin{array}{c}
\nu _{i} \\ 
e_{i}%
\end{array}%
\right) _{L}$ and the right chirality by means of a singlet, $E_{Ri}=e_{iR}$  and  $\Phi _{k}$ are the two Higgs doublets, moreover three complex dimensionless matrices (in the flavor space)  $3\times 3$ no-diagonal
$\eta _{ij}^{E},\eta _{ij}^{U},\eta _{ij}^{D}$ y $\xi _{ij}^{E},\xi _{ij}^{U},\xi _{ij}^{D}$ are associated to each doublet. These matrices correspond to the coupling constants of Yukawa between the fermionic field and the two Higgs fields.
The superindice $^{0}$ indicates that the fermion fields are not  eigenstates of mass yet  and the overlined signs show that constants of Yukawa and the Higgs doublets have been rotated. Finally, the extension of the Lagrangian to obtain Dirac's mass terms takes the form of,
%---------------------------------------------------------------------------------------------------------------------------------------------------
%\overline{L_{Li}'}\eta_{ij}^{E}\Phi_{1}E_{Rj}'+\overline{L_{Li}'}\xi_{ij}^{E}\Phi_{2}E_{Rj}'+de la ecuacion siguiente
\begin{eqnarray}
-\mathcal{L}_{Y}^{Ext-l} = \overline{L_{Li}^{0}}\eta_{ij}^{N}\widetilde{\Phi_{1}}N_{Rj}^{0}+\overline{L_{Li}^{0}}\xi_{ij}^{N}\widetilde{\Phi_{2}}N_{Rj}^{0}+h.c,.\label{ext}
\end{eqnarray}

\section{See-Saw Mechanims SSM}
\subsection{SSM type I}
\noindent It has been said that there are three generations for neutrinos. Then extending the term of Dirac-Majorana in the case that there are $n$-generations, there will be $n$-active fields
 $\nu_{\alpha_{i} L}$ \footnote{$\alpha_n$ 
 corresponding flavors} and then we must include
$n$-sterile fields $\nu_{sR}$.  Thus, we will add a fermion singlet for each generation,%------------------------------------------------------------------------------------------------------------------------------------------------
\begin{eqnarray}
 \mathcal{L}^{M+D}=\mathcal{L}^{ML}+\mathcal{L}^{MR}+\mathcal{L}^{D},
\end{eqnarray}
\noindent con
\begin{eqnarray*}
 \mathcal{L}^{ML}=\frac{1}{2}\sum_{i,j=1}^{n}\nu_{\alpha_{i}L}^{T}\widehat{C}^{\dagger}M_{L}^{\alpha_{i}\alpha_{j}}\nu_{\alpha_{j}L}+h.c\hspace{3em}
\mathcal{L}^{MR}=\frac{1}{2}\sum_{s,s'=1}^{n}\nu_{sR}^{T}\widehat{C}^{\dagger}M_{R}^{ss'}\nu_{s'R}+h.c\end{eqnarray*}

\begin{eqnarray*}
\mathcal{L}^{D}=-\sum_{j}\sum_{s}\overline{\nu_{sR}}M_{R}^{s\alpha_{j}}\nu_{\alpha_{j}L}+h.c,
\end{eqnarray*}
%--------------------------------------------------------------------------------------------------------------------------------------------------
\noindent $M_{L},M_{R},M_{D}$ are matrices of $n\times n$ and they will be complex one, with symmetrical $M_{L},M_{R}$ .
%--------------------------------------------------------------------------------------------------------------------------------------------------
\noindent For $n$-generations, the lagrangian that generates the mass term can be written as,
%--------------------------------------------------------------------------------------------------------------------------------------------------
\begin{equation}
-\mathcal{L}^{M}= \overline{{\bf E_{L}'}}{\bf M^{E}}{\bf E_{R}}'+\frac{1}{2}{\bf \mathcal{N}^{'T}}\widehat{C}^{\dagger}{\bf M_{\nu}}{\bf \mathcal{N}'_{L}}+h.c,.
\end{equation}
\noindent As ${\bf M}_L$, is prohibited by the symmetries of the SM, while ${\bf M}_R$, it is generated on a scale of symmetry rupture of the theory of the great unification GUT.
%--------------------------------------------------------------------------------------------------------------------------------------------------
\noindent The SSM type I leads to \cite{3}
%--------------------------------------------------------------------------------------------------------------------------------------------------
\begin{eqnarray}
 &&{\bf M_{\nu}}={\bf M_{lig}}\approx-{\bf M_{D}^{T}}{\bf M_{R}^{-1}}{\bf M_{D}}\\
&&{\bf M_{Pes}}\approx {\bf M_{R}}
\end{eqnarray}
%--------------------------------------------------------------------------------------------------------------------------------------------------
\noindent Where $M_{\nu}$, is Majorana's effective mass matrix of the neutrino $\nu_{L}$; $M_R$ is the Majorana mass matrix of the neutrino $\nu_{R}$, and $M_{D}$ is Dirac's mass matrix that connects $\nu_{R}$ to $\nu_{L}$. this last one was generated via Yukawa couplings.
%donde $\eta_{ij},\xi_{ij}$, , y $\widetilde{\Phi_{2}}\equiv i\tau_{2}\Phi_{2}^{*}$,
As $M_R$ is an invariant parameter under $SU(2)\times U(1)$, there is no ligature onit. Lastly,  as $M_D$ as $M_L$, are originated only after the SSB.

\subsection{SSM type III}
\noindent We added an extra triplet of fermions under $SU(2)$, for each generation to SM \cite{4}, %, denotamos en coordenadas cartesianas como [4]
%--------------------------------------------------------------------------------------------------------------------------------------
%\begin{eqnarray} 
%&&\vec{\Sigma}'_{Ri}=\left(\Sigma_{R}^{'1},\Sigma_{R}^{'2},\Sigma_{R}^{'3}\right)\\
%&\vec{\Sigma^{C}}'_{Ri}=\left(\Sigma_{R}^{'1c},\Sigma_{R}^{'2c},\Sigma_{R}^{'3c}\right),
%\end{eqnarray}
%--------------------------------------------------------------------------------------------------------------------------------------
%\noindent donde $i=1,2,3$, es necesario tener en cuenta que los campos $\Sigma_{1},\Sigma_{2}$, no son estados propios del operador de carga el\'{e}ctrica, 
%por lo cuales se escribir\'{a}n en la representaci\'{o}n adjunta del grupo de isoesp\'{i}n d\'{e}bil $SU(2)$ con hipercarga ($Y=0$)\cite{readj}.
\noindent The part of the Lagrangian responsible for the mass of the leptons can be written as
%--------------------------------------------------------------------------------------------------------------------------------------
\begin{equation}
 -\mathcal{L}=\left[\eta_{ij}^{E}\overline{E'_{Ri}}\Phi_{1}^{\dagger}L'_{Lj}+\sqrt{2}\Phi_{2}^{C\dagger}\overline{\Sigma'_{Ri}}\eta_{ij}^{\Sigma}L'_{Lj}
 +h.c\right]+\frac{1}{2}M_{ij}^{\Sigma}\left[\overline{\Sigma_{Ri}}\Sigma_{Rj}^{C}+h.c\right],
\end{equation}
%--------------------------------------------------------------------------------------------------------------------------------------
\noindent where $E_{R}$, $L_{L}$, are the usual components of right and left chirality for leptons respectively and $\eta_{ij}^{\Sigma},
\eta_{ij}^{E}$ are the $3\times3$  matrices of coupling constants of Yukawa. After the SSB, for the case of a generation and when analyzing only the part in which the mass term is generated, we have
%--------------------------------------------------------------------------------------------------------------------------------------
%\begin{equation}
%-\mathcal{L}=\frac{\eta_{ji}^{E}v_{1}}{\sqrt{2}}\overline{e'_{Ri}}e'_{Lj}+\frac{\eta_{ij}^{\Sigma}v_{2}}{\sqrt{2}}\overline{\Sigma_{Ri}^{'0}}\nu'_{Lj}+
%\eta_{ij}^{\Sigma}v_{2}\overline{\Sigma_{Ri}^{'-}}e'_{Lj}+M_{ij}^{\Sigma}\overline{\Psi'}_{Ri}\Psi'_{Lj}+
%\frac{M_{ij}^{\Sigma}}{2}\overline{\Sigma_{Ri}^{'0}}\Sigma_{Rj}^{'0}+h.c, 
%\end{equation}
%--------------------------------------------------------------------------------------------------------------------------------------
%\noindent con
%--------------------------------------------------------------------------------------------------------------------------------------
%\begin{eqnarray}
%\left[\overline{\Sigma_{Ri}^{'-}}\Sigma_{Rj}^{'+c}+\overline{\Sigma_{Ri}^{'+c}}\Sigma_{Rj}^{'-}\right]=\left[\overline{\Psi'}_{Ri}\Psi'_{Lj}+
%\overline{\Psi'_{Li}}\Psi'_{Rj}\right]=\overline{\Psi}'_{Ri}\Psi_{Lj}+h.c,
%\end{eqnarray}
%--------------------------------------------------------------------------------------------------------------------------------------
\noindent the following mass matrix for the neutrino,
\begin{eqnarray}
-\mathcal{L}^{\nu}=\frac{1}{2}\left(\begin{array}{cc}
\overline{\nu_{Li}^{'c}} & \overline{\Sigma_{Ri}^{'0}}\end{array}\right)\left(\begin{array}{cc}
0 & v_{2}\left(\eta_{ij}^{\Sigma}\right)^{T}/\sqrt{2}\\
v_{2}\eta_{ij}^{\Sigma}/\sqrt{2} & M_{ij}^{\Sigma}\end{array}\right)\left(\begin{array}{c}
\nu'_{Lj}\\
\Sigma_{Rj}^{'0c}\end{array}\right)+h.c,\label{trine}
\end{eqnarray}
\noindent By imposing the symmetry $Z_{2}$, the neutrino mass depends only on one of the VEV $v_{2}$, while in the case of the mass matrix for the charged leptons. Both doublets are coupled to give mass to the leptons of SM. The value of $v_{2}$ is determined by the mass scale of the neutrino and is independent of the mass scale of the other fermions. Therefore, the values of the neutrino mass can be small without reducing the value of the coupling constants of Yukawa.

\noindent The SSM Type III leads to neutral leptons,
%--------------------------------------------------------------------------------------------------------------------------------------
\begin{eqnarray}
{\bf M_{Lig}}\approx-{\bf M_{D}^{T}}{\bf M_{\Sigma}^{-1}}{\bf M_{D}}\hspace{1cm}
{\bf M_{Pes}}\approx {\bf M_{\Sigma}},
\end{eqnarray}
%--------------------------------------------------------------------------------------------------------------------------------------
\noindent With
%--------------------------------------------------------------------------------------------------------------------------------------
\begin{equation}
 {\bf M_{D}}=\frac{v_{2}{\bf \eta_{ij}^{\Sigma}}}{\sqrt{2}},
\end{equation}
%--------------------------------------------------------------------------------------------------------------------------------------
\noindent and for the charged leptons 
%--------------------------------------------------------------------------------------------------------------------------------------
\begin{eqnarray}
 {\bf M_{L}^{Diag}}&\approx&\frac{v_{1}}{\sqrt{2}}{\bf \eta_{ij}^{E}}\\
 {\bf M_{\Sigma}^{Diag}}&\approx& {\bf M_{\Sigma}}-{\bf M_{DT}}{\bf M_{\Sigma}^{-1}}{\bf M_{DT}},
\end{eqnarray}
%--------------------------------------------------------------------------------------------------------------------------------------
\noindent Since, $({\bf m_{DT}})_{kj}\ll({\bf M^{\Sigma}})_{kj}$, at first order in the expansion\begin{eqnarray}
{\bf M_{\Sigma}^{Diag}}={\bf M_{\Sigma}},
\end{eqnarray}
%--------------------------------------------------------------------------------------------------------------------------------------
\noindent with
%--------------------------------------------------------------------------------------------------------------------------------------
\begin{eqnarray}
 {\bf M_{DT}}=v_{2}{\bf \eta_{ij}^{\Sigma}}\hspace{0,5cm}{\bf M_{L}}=\frac{v_{1}}{\sqrt{2}}{\bf \eta_{ij}^{E}}.
\end{eqnarray}
%-------------------------------

\section{Singlet and Triplet of Majorana}
\noindent The term of Yukawa for the leptonic sector before the spontaneous break of symmetry in the case of a generation will correspond to
%--------------------------------------------------------------------------------------------------------------------------------------
\begin{eqnarray}
 -\mathcal{L}^{l}&=&\overline{L_{Li}'}\eta_{ij}^{N}\widetilde{\Phi_{1}}N_{Rj}'+\overline{L_{Li}'}\xi_{ij}^{E}\widetilde{\Phi_{2}}N_{Rj}'\notag\\
&+&\left[\eta_{ij}^{E}\overline{E'_{Ri}}\Phi_{1}^{\dagger}L'_{Lj}+\sqrt{2}\Phi_{2}^{C\dagger}\overline{\Sigma'_{Ri}}\eta_{ij}^{\Sigma}L'_{Lj}+h.c\right]
+\frac{1}{2}M_{ij}^{\Sigma}\left[\overline{\Sigma'_{Ri}}\Sigma_{Rj}^{'C}+h.c\right]+h.c,\notag\\
\end{eqnarray}
%--------------------------------------------------------------------------------------------------------------------------------------
\noindent After the SSB, the part of the Lagrangian involved in the mass of the neutrinos will be%--------------------------------------------------------------------------------------------------------------------------------------
%\begin{equation}
% \mathcal{L}^{L}=\mathcal{L}^{LC}+\mathcal{L}^{LN},
%\end{equation}
%--------------------------------------------------------------------------------------------------------------------------------------
%\noindent Siendo
%--------------------------------------------------------------------------------------------------------------------------------------
%\begin{eqnarray}
%-\mathcal{L}^{LC}&=&\left(\frac{\eta_{ij}^{e}v_{1}+\xi_{ij}^{e}v_{2}}{\sqrt{2}}\right)\overline{e_{L}^{'}}e_{R}^{'}
% +\eta_{ij}^{\Sigma}v_{2}\overline{\Sigma_{Ri}^{'-}}e'_{Lj}+M_{ij}^{\Sigma}\overline{\Psi'}_{Ri}\Psi'_{Lj}+h.c,\notag\\
%&=&\overline{e_{L}^{'}}M^{E}e_{R}^{'}+\eta_{ij}^{\Sigma}v_{2}\overline{\Sigma_{Ri}^{'-}}e'_{Lj}+M_{ij}^{\Sigma}\overline{\Psi'}_{Ri}\Psi'_{Lj}+h.c,
%\end{eqnarray}
%--------------------------------------------------------------------------------------------------------------------------------------
\begin{eqnarray}
 \mathcal{L}^{LN}&=&\frac{1}{2}m_{R}\nu_{R}^{T}\hat{C^{\dagger}}\nu_{R}-\frac{M_{ij}^{\Sigma}}{2}\overline{\Sigma_{Ri}^{'0}}\Sigma_{Rj}^{'0}
 -\left(\frac{\eta_{ij}^{N}v_{1}+\xi_{ij}^{N}v_{2}}{\sqrt{2}}\right)\overline{\nu_{L}^{'}}\nu_{R}^{'}
 -\frac{\eta_{ij}^{\Sigma}v_{2}}{\sqrt{2}}\overline{\Sigma_{Ri}^{'0}}\nu'_{Lj}+h.c,\notag\\
&=&\frac{1}{2}m_{R}\nu_{R}^{T}\hat{C^{\dagger}}\nu_{R}-\frac{M_{ij}^{\Sigma}}{2}\overline{\Sigma_{Ri}^{'0}}\Sigma_{Rj}^{'0}
-\overline{\nu_{L}^{'}}M_{ij}^{\nu}\nu_{R}^{'}-\frac{\eta_{ij}^{\Sigma}v_{2}}{\sqrt{2}}\overline{\Sigma_{Ri}^{'0}}\nu'_{Lj}+h.c,
\end{eqnarray}
%--------------------------------------------------------------------------------------------------------------------------------------
\noindent the mass term for the neutral leptons $L_{Y}^{LN}$ can be written for
\noindent the case of $3$ generations as follows
%--------------------------------------------------------------------------------------------------------------------------------------
\begin{equation}
 {\bf \mathcal{L}^{LN}}=\frac{1}{2}{\bf N_{L}^{T}}\hat{{\bf C^{\dagger}}}{\bf M_{LN}}{\bf N_{L}}+h.c,
\end{equation}
\noindent The mass matrix $9\times9$, due to the simultaneous existence of the singlet and the triplet of Majorana takes the form of:
%--------------------------------------------------------------------------------------------------------------------------------------
\begin{equation}
{\bf M_{LN}}=\left(\begin{array}{cccc}
0 & \vdots & \left(\frac{{\bf \eta_{ij}^{N}}v_{1}+{\bf \xi_{ij}^{N}}v_{2}}{\sqrt{2}}\right)^T & \left(\frac{{\bf \eta_{ij}^{\Sigma}}v_{2}}{\sqrt{2}}\right)^T\\
\cdots & \cdots & \cdots & \cdots\\
\left(\frac{{\bf \eta_{ij}^{N}}v_{1}+{\bf \xi N_{ij}^{N}}v_{2}}{\sqrt{2}}\right) & \vdots & {\bf m_{R}} & 0\\
\frac{{\bf \eta_{ij}^{\Sigma}v_{2}}}{\sqrt{2}} & \vdots & 0 & {\bf M_{\Sigma}}\end{array}\right)=\left(\begin{array}{cc}
0 & M_{D}^{T}\\
M_{D} & M_{\nu\Sigma}\end{array}\right),
\end{equation}
%--------------------------------------------------------------------------------------------------------------------------------------
\noindent with
%--------------------------------------------------------------------------------------------------------------------------------------
\begin{eqnarray}
{\bf M_{D}}\equiv\left(\begin{array}{c}
\left(\frac{{\bf \eta_{ij}^{N}}v_{1}+{\bf \xi N_{ij}^{N}}v_{2}}{\sqrt{2}}\right)\\
\frac{{\bf \eta_{ij}^{\Sigma}v_{2}}}{\sqrt{2}}\end{array}\right),\hspace{1em}\hspace{1em}\hspace{1em}\hspace{1em}\hspace{1em}
{\bf M_{\nu\Sigma}}\equiv\left(\begin{array}{cc}
{\bf m_{R}} & 0\\
0 & {\bf M_{\Sigma}}\end{array}\right),
\end{eqnarray}
%--------------------------------------------------------------------------------------------------------------------------------------
\noindent Matrices of $6\times3$ and $6\times6$, respectively. It is evident that $M_{\nu\Sigma}$ is diagonal by blocks, it is important to highlight  that the  simultaneous introduction of the singlet and the triplet of Majorana is equivalent to the Weinberg operator in the static limit of these two fermions.

%Se podr\'{i}a pensar en una, una diagonalizaci\'{o}n por bloques de un tipo similar al utilizado cuando se consideraba singlete y triplete de manera separada,
%es decir
%--------------------------------------------------------------------------------------------------------------------------------------
%\begin{equation}
%{\bf W_{\nu}^{T}}{\bf M_{LN}}{\bf W_{\nu}}={\bf M_{\nu}^{D}}=\left(\begin{array}{cc}
%{\bf M_{lig}} & 0\\
%0 & {\bf M_{Pes}}\end{array}\right)\label{sintri},
%\end{equation}
%--------------------------------------------------------------------------------------------------------------------------------------
%\noindent donde ${\bf M_{Pes}}$ es una matriz $6\times6$ correspondiente a los campos de neutrino pesados. Adem\'{a}s la matriz unitaria ${\bf W_{\nu}}$ de 
%$9\times9$ se tomar\'{a} como
%--------------------------------------------------------------------------------------------------------------------------------------
%\begin{eqnarray}
% {\bf W_{\nu}}=\left(\begin{array}{cc}
%\sqrt{{\bf 1_{3\times3}}-{\bf B}{\bf B^{\dagger}}} & {\bf B}\\
%-{\bf B^{\dagger}} & \sqrt{{\bf 1_{6\times6}}-{\bf B^{\dagger}}{\bf B}}\end{array}\right),
%\end{eqnarray}
%--------------------------------------------------------------------------------------------------------------------------------------
\section*{\bf\small 3.1 See-Saw Mechanims Singlet and Triplets of Majorana}
\noindent The SSM type I+III leads to:
%--------------------------------------------------------------------------------------------------------------------------------------
\begin{eqnarray}
 {\bf M_{Lig}}&\approx&-\left(\frac{{\bf \eta_{ij}^{N}}v_{1}+{\bf \xi_{ij}^{N}}v_{2}}{\sqrt{2}}\right)^{T}{\bf M_{R}^{-1}}\left(\frac{{\bf \eta_{ij}^{N}}v_{1}
 +{\bf \xi_{ij}^{N}}v_{2}}{\sqrt{2}}\right)-\left(\frac{{\bf \eta_{ij}^{\Sigma}}v_{2}}{\sqrt{2}}\right)^{T}{\bf M_{\Sigma}^{-1}}
 \left(\frac{{\bf \eta_{ij}^{\Sigma}}v_{2}}{\sqrt{2}}\right),\notag\\
\end{eqnarray}
\noindent The mass matrix of the light neutrinos is equal to the sum of the matrices of mass obtained in the study of the singlet and triplet separately.
Now the matrix of the heavy fields are given by
%--------------------------------------------------------------------------------------------------------------------------------------
\begin{equation}
 {\bf M_{Pes}} \approx\left(\begin{array}{cc}
{\bf M_{R}} & 0\\
0 & {\bf M_{\Sigma}}\end{array}\right),
\end{equation}
%--------------------------------------------------------------------------------------------------------------------------------------
\noindent It is evident that heavy fields are decoupled to the first order, which is a good approximation.

\section{Conclusions}
\noindent We have found that the MSS explains the small neutrino mass  due to
that the singlet has a very large mass. It implies that the masses of the left chirality neutrinos are very small compared to the  other fermions of the same family. At the same time adding a second Higgs doublet $SU (2)$ in the 2HDM-III explanains the small mass of the neutrinos in the context of the SSM type I. Additionally, we have considered the hybrid scenario  which is an extension of the SM with the simultaneous inclusion of a singlet and a triplet of Majorana . We have directly studied the SSM for this case showing that the mass matrix of the light neutrinos corresponds to the sum of the matrices obtained for each of the cases separately.

\section*{acknowledgment}

This work was supported by the physics departament of Universidad Pedag\'{o}gica y Tecnol\'{o}gica de Colombia and the fields and partciles group of the Universidad Nacional de Colombia.

\appendix

%\section{app 1}

\end{document}